\documentclass[a4paper,11pt]{article}
\usepackage{pos}
\usepackage{amsmath}
\usepackage{dsfont}
\usepackage{bbm}
\usepackage{cleveref}
\usepackage{subcaption}
\usepackage[utf8]{inputenc}
\usepackage{mathrsfs}
\usepackage{amsfonts}

\let\OLDthebibliography\thebibliography
\renewcommand\thebibliography[1]{
  \OLDthebibliography{#1}
  \setlength{\parskip}{0pt}
  \setlength{\itemsep}{0pt plus 0.3ex}
}

\DeclareMathOperator{\tr}{tr}

\newcommand{\fm}{\text{fm}}
\newcommand{\dif}{\mathrm{d}}

\newcommand{\flow}{\mathrm{fl}}

\newcommand{\chihat}{\hat\chi}
\newcommand{\Ohat}{\hat{O}}
\newcommand{\rhat}{\hat r}
\newcommand{\ahat}{\hat a}
\newcommand{\tflow}{t_{\flow}}

\title{Gauge field smearing and controlled continuum extrapolations}
\ShortTitle{Gauge field smearing and controlled continuum extrapolations}

\author*[a,b]{Andreas Risch}

\affiliation[a]{Department of Physics, University of Wuppertal, Gaussstr. 20, 42119 Wuppertal, Germany}
\affiliation[b]{John von Neumann-Institut f{\"u}r Computing NIC, Deutsches Elektronen-Synchrotron DESY,\\
Platanenallee 6, 15738 Zeuthen, Germany}

\emailAdd{andreas.risch@uni-wuppertal.de}

\abstract{When designing lattice actions, gauge field smearing is often used in the definition of the lattice Dirac operator. Too much smearing can result in uncontrolled continuum extrapolations as the short distance behaviour of the theory is mutilated, which is a situation to be avoided. As a smearing prescription we focus on the gradient flow formalism as it allows to study both smearing and physical flow simultaneously. We investigate the effect of smearing and physical flow on the scaling towards the continuum limit in pure gauge theory. We focus on the example of Creutz ratios, which provide a measure of the physical forces felt by the fermions. For suitable smearing strengths we further investigate the impact of replacing the Wilson gradient flow by stout smearing.}

\FullConference{European network for Particle physics, Lattice field theory and Extreme computing (EuroPLEx2023)\\
 11-15 September 2023\\
Berlin, Germany\\}

\begin{document}
\maketitle

\section{Introduction}

A reduction of lattice artefacts is beneficial for more reliable continuum extrapolations, in particular of short distance observables. A popular methods to alter discretisation effects is UV filtering, which is based on the application of four-dimensional gauge field smearing. The Dirac operator is evaluated on smeared gauge fields such that the action is altered into
\begin{gather}
S[U] = S_{\mathrm{g}}[U] + \overline{\Psi}\,D[\mathcal{S}[U]]\,\Psi,
\end{gather}
where $\mathcal{S}:U\mapsto \mathcal{S}[U]$ is a smearing transformation. Several smearing algorithms have been developed, e.g. HYP~\cite{Hasenfratz:2001hp}, Stout~\cite{Morningstar:2003gk}, HEX~\cite{Capitani:2006ni} and gradient flow~\cite{Narayanan:2006rf,Luscher:2010iy} smearing. Evaluating the Dirac operator on smeared gauge fields yields several advantages: The likelihood of finding small eigenvalues of $D$ is reduced, i.e. exceptional configurations can be avoided. In~\cite{Hasenfratz:2007rf} even at very coarse lattice spacings the Wilson Dirac operator defined with nHYP gauge links could be shown to exhibit a spectrum with a well-defined spectral gap. The same was shown for stout smearing in~\cite{Durr:2008rw}. This is particularly helpful for the simulation of mass non-degenerate quarks as the fermion determinant is not necessarily positive in such a scenario~\cite{Mohler:2020txx}. Gauge field smearing also has an impact on improvement coefficients and renormalisation constants. In~\cite{Hasenfratz:2007rf} it was observed that the improvement coefficient $c_{\mathrm{SW}}$ approaches its tree-level value when gauge field smearing is applied. The amount of renormalisation in $Z_{\mathrm{V}}$ is also reduced. However, the application of too much smearing may significantly alter the UV structure of the lattice theory and therefore continuum extrapolations based on data from insufficiently small lattice spacings may become unreliable. It is therefore relevant to study the range of smearing strengths that still allow for controlled continuum extrapolations. As a first step towards a smeared action setup with fermions we study smeared observables
\begin{gather}
\langle O_{\mathcal{S}}[U] \rangle = \langle O[\mathcal{S}[U]] \rangle
\end{gather}
in pure gauge theory. We investigate the influence of smearing on continuum extrapolations of Creutz ratios~\cite{Creutz:1980wj}, which provide a measure of the physical forces felt by the fermions caused by the gauge field. For a previous account of this effort we refer the reader to~\cite{Risch:2022vof,Risch:2023qjy}.

\section{The gradient flow formalism, gradient flow smearing and physical gradient flow}
\label{sec:gradientflow}

In this work we focus on the gradient flow formalism~\cite{Luscher:2010iy} as a smearing procedure. We start from the continuum four-dimensional Yang-Mills action $S_{\mathrm{YM}}=-\frac{1}{2g_0^2}\int \dif^4 x \tr( F_{\mu\nu}(x)F_{\mu\nu}(x))$. $F_{\mu\nu} = \partial_{\mu}A_{\nu}-\partial_{\nu}A_{\mu}+[A_{\mu},A_{\nu}]$ denotes the field strength tensor and $A_{\mu}(x)$ the corresponding gauge field. In the gradient flow formalism a gauge field $B_{\mu}(x,\tflow)$ is introduced, where $\tflow\geq 0$ is the so called gradient flow time. At $\tflow=0$ the standard gauge field $A_{\mu}(x)$ is used as an initial condition for the flow time evolution, i.e. $B_{\mu}(x,0) = A_{\mu}(x)$. The evolution is then governed by the gauge-covariant flow equation
\begin{gather}
\frac{\partial}{\partial \tflow}B_{\mu}(x,\tflow) = -\frac{\delta S_{\mathrm{YM}}[B]}{\delta B_{\mu}(x,\tflow)} = D_{\nu}G_{\nu\mu}(x,\tflow),
\end{gather}
where $G_{\mu\nu} = \partial_{\mu}B_{\nu}-\partial_{\nu}B_{\mu}+[B_{\mu},B_{\nu}]$ denotes the generalised field strength tensor and $D_{\mu} = \partial_{\mu} + [B_{\mu}, \cdot]$ the generalised covariant derivative. Performing a leading-order perturbative expansion it was shown that the gauge field $B_{\mu}(x,\tflow)$ is a spherically smoothed version of $A_{\mu}(x)$ with mean-square radius $r_{\mathrm{sm}} = \sqrt{8 \tflow}$~\cite{Luscher:2010iy}, i.e. in the direction of positive flow time the gradient flow possesses a smoothing property. In~\cite{Luscher:2011bx} it was shown perturbatively to all loop orders that any functional of the flowed fields $B_{\mu}(x,\tflow)$ at strictly positive $\tflow$ is finite, assuming that the four-dimensional theory has been renormalised. Consequently, no additional renormalisation has to be applied. The Wilson gradient flow~\cite{Luscher:2010iy} is used as a lattice discretisation of the Yang-Mills gradient flow. The flow equation is then integrated numerically using an explicit 3rd-order Runge-Kutta integration scheme~\cite{Luscher:2010iy} with a step size $\frac{\Delta \tflow}{a^{2}}$ never exceeding $0.01$.

The gradient flow will be applied to the gauge field in two scenarios: In the first scenario, which we refer to as gradient flow smearing, the gradient flow time and consequently the smearing radius vanishes in the continuum limit. Hence the continuum theory is unaltered. This can be achieved by fixing the gradient flow time in lattice units, i.e. $\frac{8\tflow}{a^{2}} = \mathrm{const}$. The second scenario, in which the flow time is fixed in physical units, i.e. $\tflow / t_0= \mathrm{const}$, we refer to as a physical gradient flow. In principle, $t_0$ may be any physical scale of the theory. We make use of the reference flow time introduced in~\cite{Luscher:2010iy}, which we will define in \cref{sec:latticesetup}. In this scenario the continuum theory is altered. This type of alteration of an observable's continuum limit can also be understood as a modification of the definition of the observable itself, i.e. the physical gradient flow allows to construct new observables.

\section{Combined continuum extrapolation and small flow time expansion}
\label{sec:combinedextrapolation}

In the following we consider a dimensionless observable $\Ohat$, which does not require a renormalisation and hence is finite in the continuum limit. We will understand this observable as a function of the dimensionless lattice spacing parameter $\ahat\equiv\frac{a}{\sqrt{8t_{0}}}$ and the flow time parameter $\varepsilon=\frac{\tflow}{t_{0}}$. Due to the finiteness of the observable the continuum limit and the zero flow time limit can be interchanged, i.e. $\lim_{\ahat \rightarrow 0}\lim_{\varepsilon \rightarrow 0} \Ohat = \lim_{\varepsilon \rightarrow 0}\lim_{\ahat \rightarrow 0} \Ohat$. In this case the two scenarios discussed in \cref{sec:gradientflow} have a common limit where both $a=0$ and $t_\flow=0$. Therefore, a combined Symanzik and small flow time expansion is possible and well-defined. The double expansion of the observable reads
\begin{align}
\Ohat &= \sum_{i,j \geq 0}c_{ij} \ahat^{i}\varepsilon^{j}.\label{eq:obsdoubleexpansionflow}
\end{align}
We neglect logarithmic effects both in the lattice spacing~\cite{Husung:2021mfl} and in the flow time~\cite{Luscher:2010iy} as this investigation has only intermediate precision. Evaluating this expression in the continuum $\ahat=0$, it becomes obvious that the observable's continuum limit $\Ohat = c_{00} + \sum_{j>0}^{n}c_{0j} \varepsilon^{j}$ can be altered by a physical gradient flow. $c_{00}$ denotes the continuum limit at vanishing flow time. In this work, we are primarily interested in the effect of smearing on the continuum extrapolation. To demonstrate that \cref{eq:obsdoubleexpansionflow} also describes the observable's lattice spacing dependence at fixed smearing strength $\frac{8\tflow}{a^{2}}$, we observe that the latter is parametrised by $\frac{\varepsilon}{\ahat^{2}}=\frac{8\tflow}{a^{2}}$. The expansion can therefore we rewritten as a function of the lattice spacing and the smearing strength:
\begin{align}
\Ohat &= \sum_{i,j \geq 0}c_{ij} \ahat^{i+2j}\Big(\frac{\varepsilon}{\ahat^{2}}\Big)^{j} = \sum_{i,j \geq 0}c_{ij} \ahat^{i+2j}\Big(\frac{8\tflow}{a^{2}}\Big)^{j}. \label{eq:obsdoubleexpansionsmearing}
\end{align}
Evaluating the smearing expansion in the continuum limit $\ahat=0$ yields $\Ohat = c_{00}$, i.e. the continuum limit is independent of the smearing strength by construction. The main advantage of this combined Symanzik and small flow time expansion is that data measured at various small $\ahat\equiv\frac{a}{\sqrt{8t_{0}}}$ and $\varepsilon=\frac{\tflow}{t_{0}}$ can be combined to determine the coefficients $c_{ij}$, from which the lattice spacing dependence can be reconstructed for any sufficiently small smearing strength or flow time parameter.

\section{Lattice setup}
\label{sec:latticesetup}

\begin{table}[h]
\begin{center}
\begin{tabular}{|l|lll|ll|l|}
\hline
ensemble & $\beta$ & $T/a$ & $L/a$ & $a\,[\fm]$ & $L\,[\fm]$ & $t_{0}/a^{2}$ \\
\hline
sft1 & 6.0662 & 80 & 24 & 0.0820(5) & 1.968(12) & 3.990(9) \\
sft2 & 6.2556 & 96 & 32 & 0.0616(4) & 1.971(12) & 7.070(17) \\
sft3 & 6.5619 & 96 & 48 & 0.04031(26) & 1.935(12) & 16.52(6)\\
sft4 & 6.7859 & 192 & 64 & 0.03010(19) & 1.927(12) & 29.60(10)\\
sft5 & 7.1146 & 320 & 96 & 0.01987(13) & 1.908(12) & 67.94(23)\\
\hline
\end{tabular}
\caption{Parameters of the SU(3) gauge ensembles~\cite{Husung:2017qjz} and computed reference flow time $t_{0}/a^{2}$ in lattice units.}
\label{tbl:ensembles}
\end{center}
\end{table}

This study is based on $\mathrm{SU}(3)$ Yang Mills theory gauge ensembles~\cite{Husung:2017qjz} using the Wilson plaquette action, where temporal open boundary conditions~\cite{Luscher:2011kk} are imposed to alleviate topology freezing. An overview of the gauge ensembles is given in \cref{tbl:ensembles}. The reference flow time $t_{0}$~\cite{Luscher:2010iy} is used as a scale to construct dimensionless quantities. To define $t_{0}$ we make use of the action density
\begin{align}
E(x,\tflow) = -\frac{1}{2}\sum_{\mu,\nu}\tr\big(G^{\mathrm{clv}}_{\mu\nu}(x,\tflow)\,G_{\mu\nu}^{\mathrm{clv}}(x,\tflow)\big),
\end{align}
where $G^{\mathrm{clv}}$ denotes the field strength tensor in the clover discretisation~\cite{Sheikholeslami:1985ij}. The reference flow time $t_{0}$ is then implicitly defined by~\cite{Luscher:2010iy}
\begin{align}
t^{2}_{0}\,\langle E(x,t_0) \rangle &= 0.3.
\end{align}
Numerical values are listed in \cref{tbl:ensembles}. The physical value of $t_{0} = 0.0268(3)\,\fm^{2}$ is obtained from the force parameter $r_{0}$~\cite{Sommer:1993ce}, where for illustration a value of $r_{0} = 0.5\,\fm$ is used. The lattice spacing varies between $0.08\,\fm$ and $0.02\,\fm$ and the spatial extent between $1.9\,\fm$ and $2\,\fm$.

\section{Creutz ratios and gradient flow}

Creutz ratios~\cite{Creutz:1980wj} are suitable observables for a study in pure gauge theory as they possess a finite continuum limit. The latter are constructed from planar rectangular Wilson loops $W(r, t) \equiv \langle \tr( P \exp(\oint_{\gamma(r,t)} \dif x_{\mu}A_{\mu}(x))) \rangle$, which are obtained from the gauge field by a path-ordered integral along a rectangular closed path $\gamma(r,t)$. In lattice gauge theory these objects are discretised as
\begin{gather}
W(r,t) = \Big\langle \tr\Big(\prod_{(x,\mu)\in\gamma(r,t)} U_{\mu}(x)\Big) \Big\rangle.
\end{gather}
Creutz ratios are obtained from Wilson loops by $
\chi(r, t) \equiv -\frac{\partial}{\partial t}\frac{\partial}{\partial r} \ln(W(r, t))$. To obtain $O(a^{2})$ lattice artefacts the latter definition is discretised making use of central differences~\cite{Okawa:2014kgi}:
\begin{gather}
\chi\Big(t+\frac{a}{2}, r+\frac{a}{2}\Big) \equiv \frac{1}{a^{2}}\ln\Big(\frac{W(t+a, r)\cdot W(t, r+a)}{W(t, r) \cdot W(t+a, r+a)}\Big). \label{eq:creutzlat}
\end{gather}
The static quark anti-quark force can be extracted in the limit of an infinite time extent, $\chi(r,t) \rightarrow F_{\mathrm{\overline{q}q}}(r)$ for $t\rightarrow\infty$~\cite{Okawa:2014kgi}.

In the following discussion we will only focus on diagonal Creutz ratios $\chi(r, t)$ with $r=t$, which we abbreviate as $\chi(r)\equiv\chi(r, r)$. We compute the latter in lattice units $(\chi\cdot a^{2})(\frac{r}{a})$ for various half integer distances $\frac{r}{a}=1.5,2.5,\ldots$ based on gauge configurations which gradient flow smearing was applied to. We use $t_0$ to define dimensionless Creutz ratios, i.e. we analyse $\chihat \equiv \chi\cdot 8 t_{0}$ as a function of $\rhat \equiv\frac{r}{\sqrt{8t_{0}}}$. In our measurements we implement the two scenarios for scaling the flow time via
\begin{equation}
\label{eq:scenarios}	
\frac{8t_\flow}{a^2} =
\begin{cases}
0,\; 0.25,\; 0.5,\; \ldots,\; 2, \; 2.5, \ldots, \;3.5,\;4,\; 5,\; 6,\; 7, \; 8 & \text{smearing}\\
\frac{8t_0}{a^2} \times0.0146 \times j\,,\;\; j \in\{0,\;1,\;\ldots ,\;4\} & \text{physical flow}.
\end{cases}
\end{equation}
The computation is based on the openQCD~\cite{Luscher:openQCD} package and utilises B. Leder's program for measuring Wilson loops~\cite{Leder:wloop,Donnellan:2010mx}. For the data analysis the python3 package pyobs~\cite{Bruno:pyobs} is used, which implements the $\Gamma$-method~\cite{Wolff:2003sm} for Monte Carlo error estimation.

\begin{figure}
\centering
\includegraphics[width=0.495\textwidth]{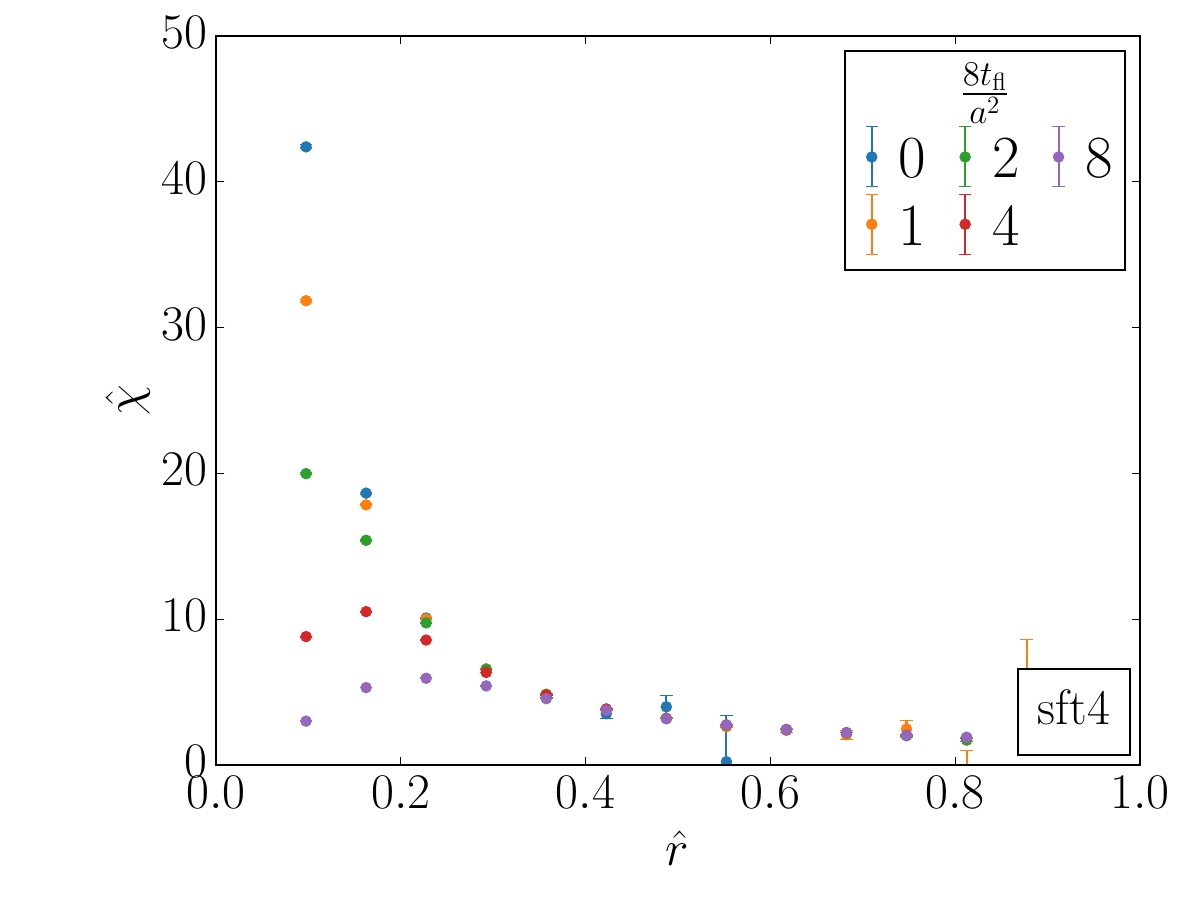}
\includegraphics[width=0.495\textwidth]{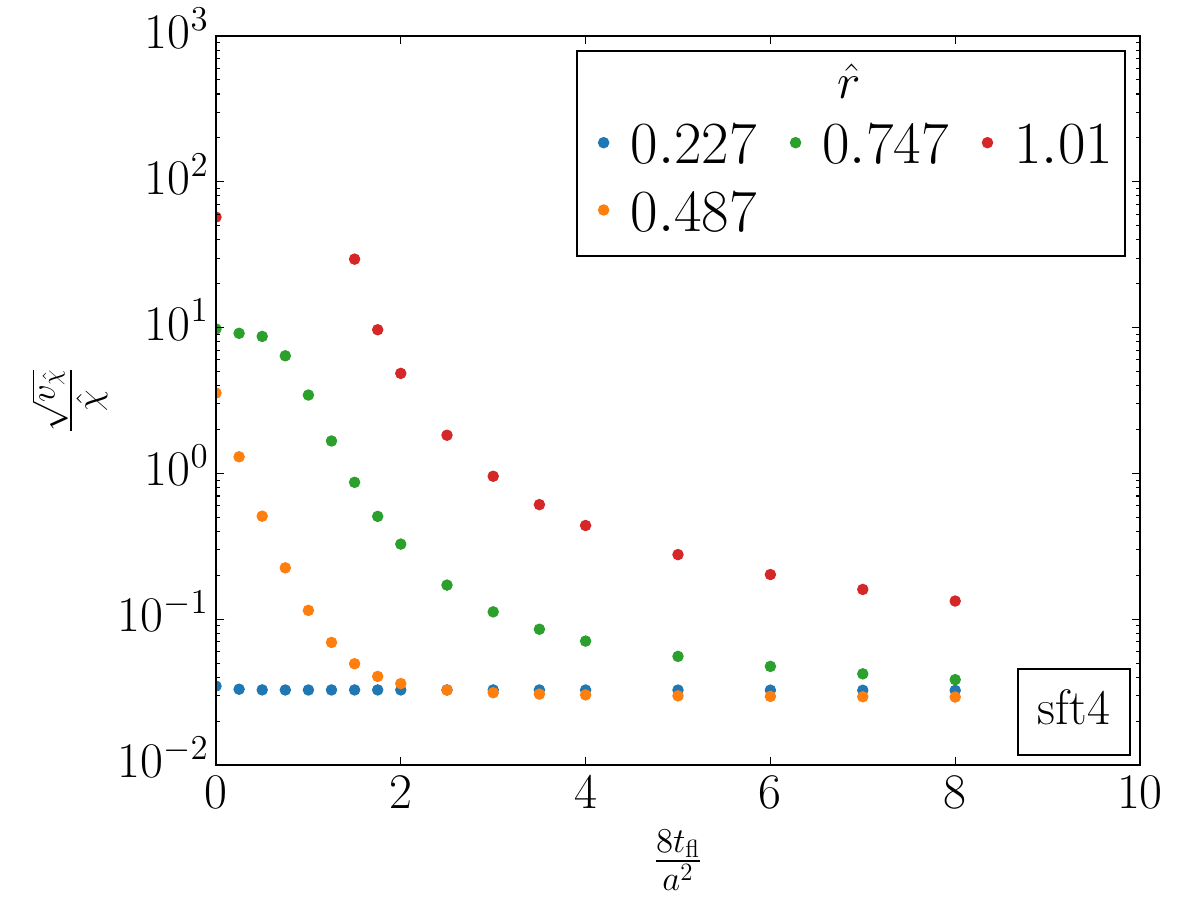}
\caption{Dimensionless Creutz ratio $\hat\chi$ and relative variance $\frac{\sqrt{v_{\hat\chi}}}{\hat\chi}$ as functions of the flow time $\frac{8t_{\flow}}{a^{2}}$ and the distance $\hat r$ on the ensemble sft4.\label{fig:smearing}}
\end{figure}

As discussed in the introduction smearing is commonly used to reduce UV fluctuations in gauge fields, which also has an impact on the variance of observables. In \cref{fig:smearing} the dimensionless diagonal Creutz ratio $\chihat$ and its relative variance $\frac{\sqrt{v_{\hat\chi}}}{\hat\chi}$ are displayed as functions of the distance $\rhat$ and the smearing strengths $\frac{8t_{\flow}}{a^{2}}$ for the ensemble sft4. We observe that the $\sim \frac{1}{r^2}$ short distance behaviour is smoothed by the gradient flow at distances $r \lessapprox \sqrt{8t_\flow}$. Consequently, the path to the continuum and hence lattice artefacts are altered in the smearing scenario. This effect becomes smaller at larger distances where the smearing has less impact. We observe that the relative variance of the Creutz ratio $\frac{\sqrt{v_{\hat\chi}}}{\hat\chi}$ grows with growing distances. Applying gradient flow smearing the relative variance shrinks with growing flow time at all distances~\cite{Okawa:2014kgi}. However, smearing the gauge fields does not lead to an arbitrary large reduction of the relative variance, which seems to be almost independent of the distance $\rhat$.

\section{Interpolation of Creutz ratios}

\begin{figure}
\centering
\includegraphics[width=0.495\textwidth]{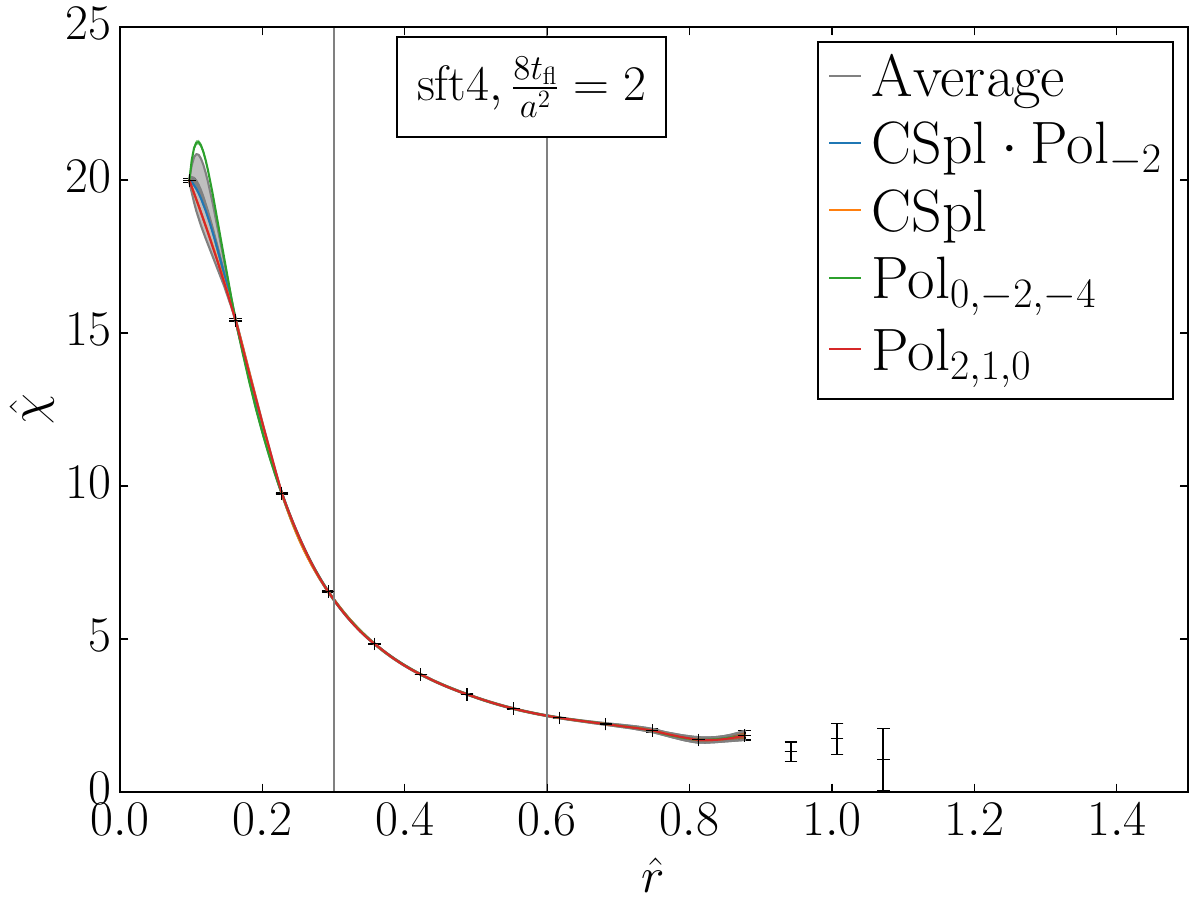}
\includegraphics[width=0.495\textwidth]{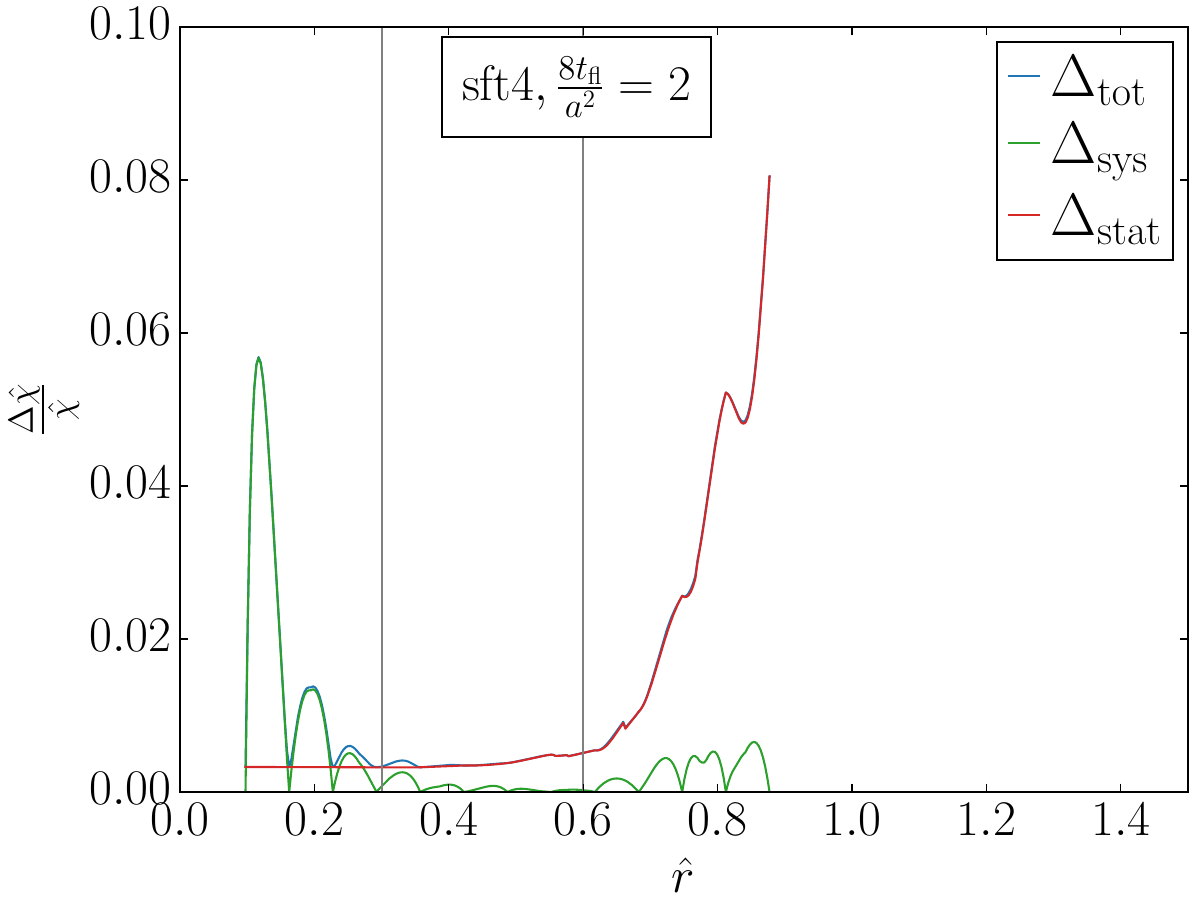}
\caption{Dimensionless Creutz ratio $\chihat$ as a function of the distance $\rhat$ on the ensemble sft4 with a gradient flow time $\frac{8 \tflow}{a^{2}}=2$ for various interpolation models.}
\label{fig:interpolation}
\end{figure}

In order to perform a continuum extrapolation of a Creutz ratio $\chihat(\rhat)$ evaluated at a fixed distance $\rhat$, $\chihat(\rhat)$ has to be known on all ensembles at $\rhat$. Therefore, an interpolation of $\chi\cdot a^{2}$ as a function of $\frac{r}{a}$ is applied. This interpolation is however not unique but depends on the specific interpolation model. In \cref{fig:interpolation} the results of several interpolation models are shown for a smearing strength $\frac{8 \tflow}{a^{2}}=2$ for the ensemble sft4. The relative statistical, systematic and total variances are also displayed. Details on the used interpolation models are described in~\cite{Risch:2022vof}. The interpolation models significantly differ at short distances, where the systematic error associated with the choice of the interpolation model dominates the overall error. The interpolations also differ at larger distances where the data fluctuates due to the loss of the signal. However, the dominant source of uncertainty in this region is the statistical error. In this work we focus on the region $0.3 \leq\rhat\leq 0.6$ ($0.14\,\fm \leq r\leq 0.28\,\fm$), where lattice artefacts are not uncontrollably large and the statistical as well as the systematic uncertainties for the ensembles sft1-4 are sufficiently small.

\section{Influence of the smearing strength on the continuum extrapolation}

\begin{figure}
\centering
\includegraphics[width=0.75\textwidth]{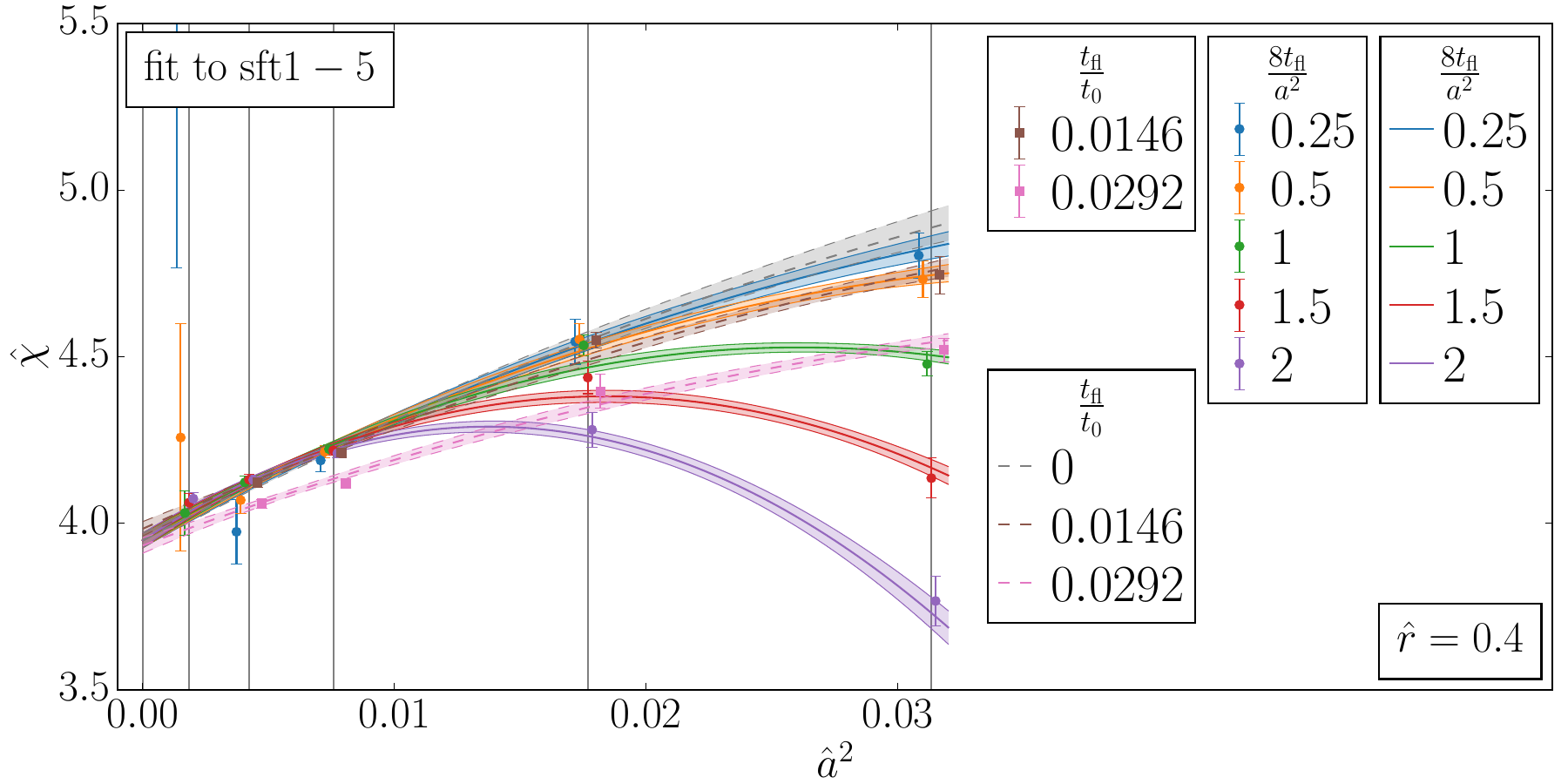} 
\\
\includegraphics[width=0.75\textwidth]{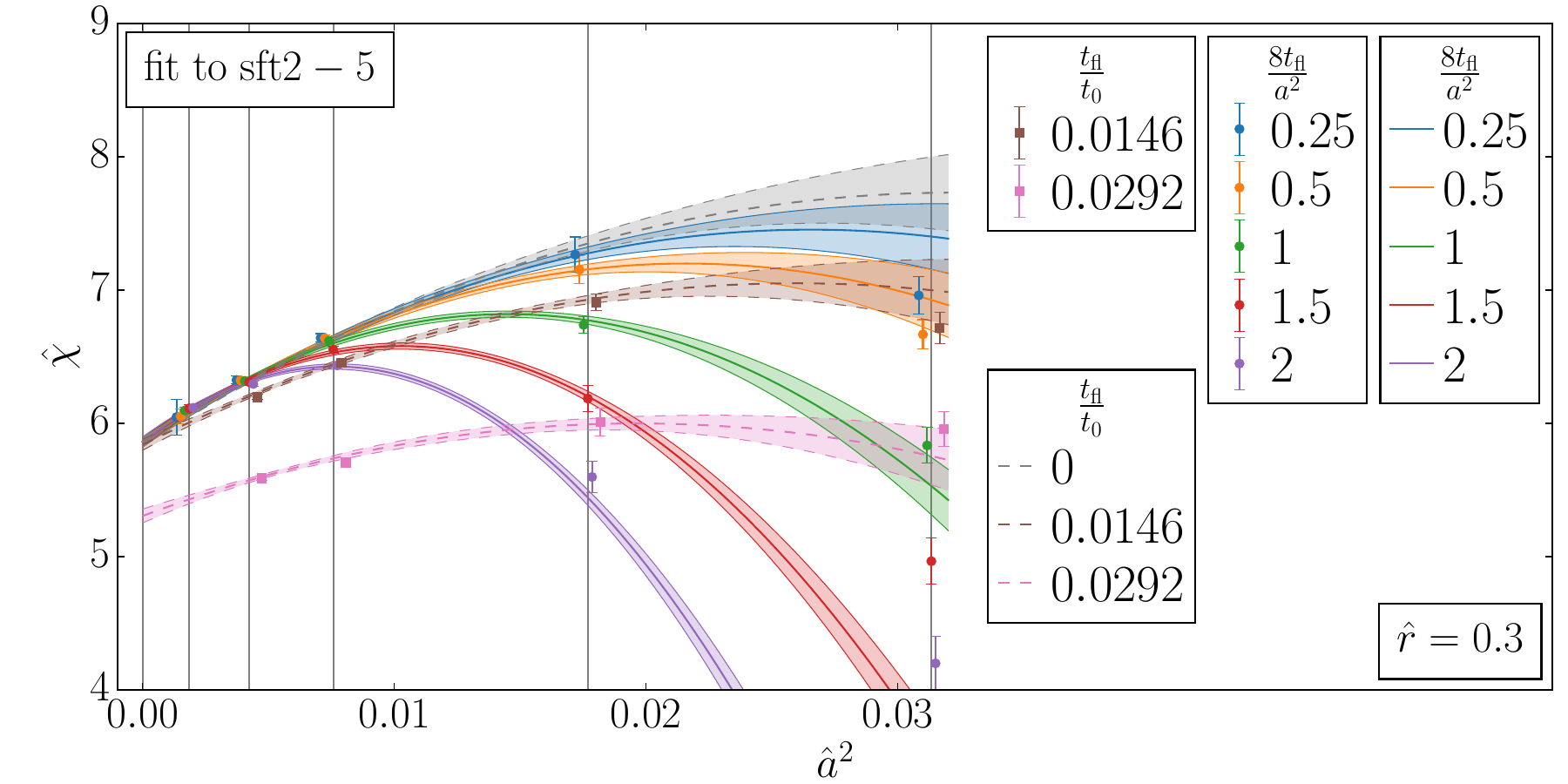}
\caption{Dimensionless Creutz ratio $\chihat\equiv\chi\cdot 8t_{0}$ as a function of the lattice spacing $\ahat\equiv\frac{a}{\sqrt{8t_{0}}}$ at a distance of $\rhat\equiv\frac{r}{\sqrt{8t_{0}}}=0.3$ (top) and $0.4$ (bottom) ($r=0.14/0.18\,\fm$). Extrapolations for several gradient flow smearing strengths $\frac{8\tflow}{a^{2}}$ (solid) and for several physical gradient flows $\frac{\tflow}{t_{0}}$ (dashed). Solid lines and circles belong to gradient flow smearing, whereas dashed lines and squares represent a physical gradient flow. Data points have been shifted for better visibility.}
\label{fig:continuum}
\end{figure}

As discussed in \cref{sec:combinedextrapolation} we perform a global continuum extrapolation at fixed distance $\rhat$ combining data from both the smearing and the physical gradient flow scenarios, c.f. \cref{eq:scenarios}. A fit ansatz for the extrapolation is obtained by truncating the expansion \cref{eq:obsdoubleexpansionflow}:
\begin{align}
\chihat_{\mathrm{tr}} &= c_{00} + c_{20} \ahat^{2} + c_{40} \ahat^{4} + c_{01}\varepsilon + c_{21}\ahat^{2}\varepsilon + c_{02}\varepsilon^{2}. \label{eq:fitflow}
\end{align}
To make the description of smearing more visible the latter fit ansatz can also be written as a truncation of \cref{eq:obsdoubleexpansionsmearing}:
\begin{align}
\chihat_{\mathrm{tr}} &= c_{00} + c_{20}\,\Big(1+\frac{c_{01}}{c_{20}}\frac{8\tflow}{a^{2}}\Big)\,\ahat^{2} + c_{40}\,\Big(1 + \frac{c_{21}}{c_{40}}\frac{8\tflow}{a^{2}} + \frac{c_{02}}{c_{40}}\Big(\frac{8\tflow}{a^{4}}\Big)^2\Big)\,\ahat^{4}. \label{eq:fitsmearing}
\end{align}
In \cref{fig:continuum} the continuum extrapolations for the distances $\rhat=0.3$ and $0.4$ for both smearing and physical flow are shown. An inclusion of data from too large smearing strengths $\frac{8\tflow}{a^{2}}$ and too large physical gradient flow times $\frac{t_{0}}{\tflow}$ led to fits with a insufficient $p$-value, which is an indication of the breakdown of the Symanzik and the small flow time expansion. In principle, this depends however on the distance $\rhat$: At larger distances larger values of $\frac{8\tflow}{a^{2}}$ and $\frac{t_{0}}{\tflow}$ can be fitted with the given low-order expansion. Extrapolations for different smearing strengths $\frac{8\tflow}{a^{2}}$ are depicted with solid lines. By construction of the fit model the continuum limit is independent of $\frac{8\tflow}{a^{2}}$ in the smearing scenario as pointed already out in \cref{sec:combinedextrapolation} and hence all continuum extrapolations share a common continuum limit. We observe that for larger $\frac{8t_{\flow}}{a^{2}}$ smearing extrapolations possess an increasingly non-monotonous behaviour. Furthermore, extrapolations for different physical gradient flow times $\varepsilon=\frac{\tflow}{t_{0}}$ are depicted with dashed lines. As already anticipated in \cref{sec:combinedextrapolation} we observe a dependence of the continuum limit on $\varepsilon$, which the fit model is able to describe by construction. This dependence is more pronounced at shorter distances. For the given range of $\varepsilon$ and $\rhat$ we observe monotonous continuum extrapolations.

\begin{figure}[h]
\centering
\includegraphics[width=0.5\textwidth]{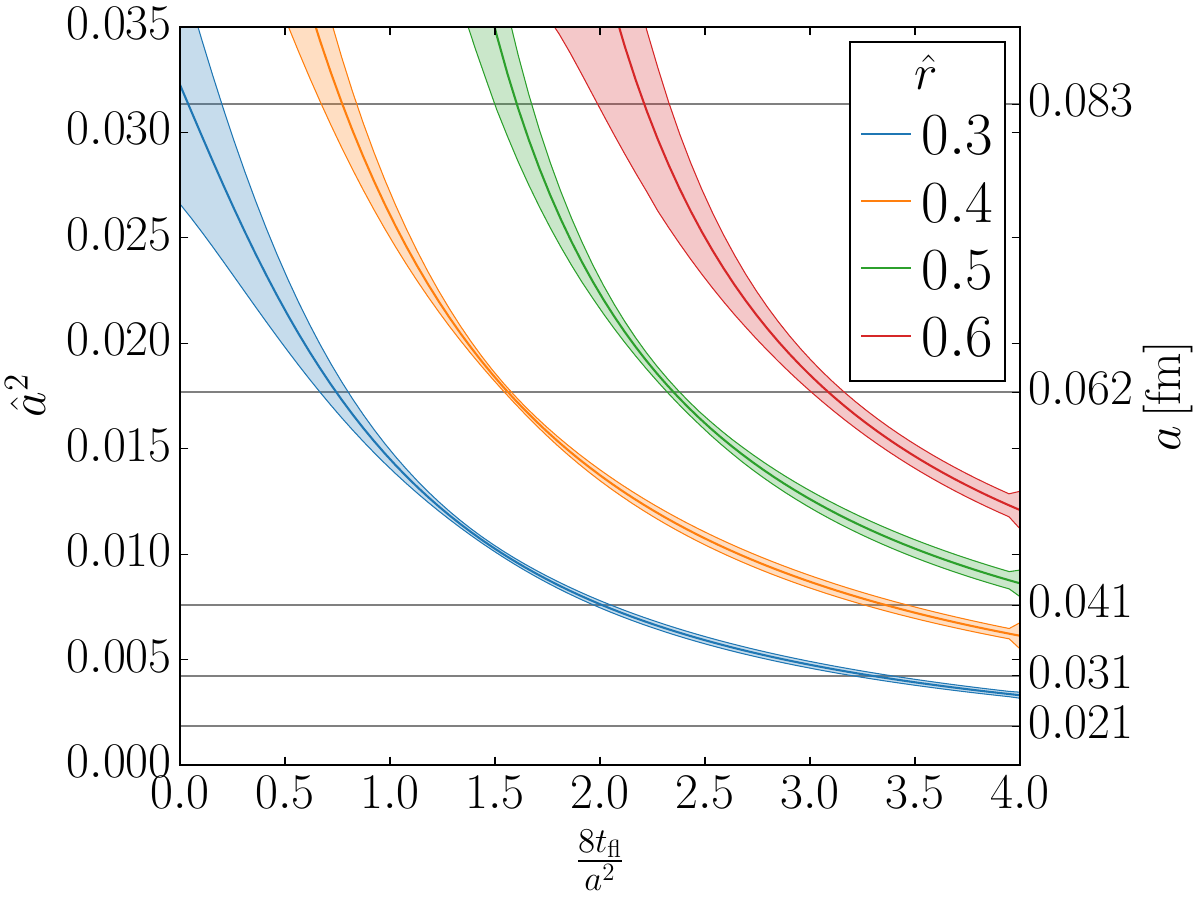}
\caption{Location of the maximum of $\chihat(\ahat)$ as a function of the smearing strength $\frac{8\tflow}{a^{2}}$ for several distances $\rhat\equiv\frac{r}{\sqrt{8t_{0}}}$.}
\label{fig:peaklocation}
\end{figure}

The requirement of a monotonous continuum extrapolation may serve as a loose criterion for a controlled continuum extrapolation, which we will apply to the smearing scenario in the following. The latter criterion will limit the smearing strength and the lattice spacing of the coarsest ensemble considered. To study monotony we track the location $\ahat_\mathrm{peak}^{2}$ of the maximum of the continuum extrapolation of $\chihat$ as a function of $\frac{8t_{\flow}}{a^{2}}$ at various distances $\rhat$. Consequently, all lattice spacings with $\ahat^{2}<\ahat_\mathrm{peak}^{2}$ belong to a continuous extrapolation. In \cref{fig:peaklocation} the location of the peak $\ahat_\mathrm{peak}^2$ is plotted as a function of the smearing radius for various distances $\rhat$. Given a minimum distance $\rhat$ that is supposed to be extrapolated reliably in the sense of a monotonous continuum extrapolation, all points $\big(\frac{8\tflow}{a^{2}},\ahat^{2}\big)$ below the curve related to the distance $\rhat$ describe to monotonous extrapolations. In particular, we observe that larger distances $\rhat$ allow for larger maximum lattice spacings and more importantly for a larger smearing strength $\frac{8\tflow}{a^{2}}$. Considering ensembles with lattice spacings $a\leq 0.06\,\fm$ and demanding for monotonous extrapolations for distances $r\geq 0.14\,\fm$ ($\rhat\geq0.3$) one should choose $\frac{8\tflow}{a^{2}}\lessapprox 1$. In order to reliably describe even smaller distances one has to consider ensembles with smaller lattice spacings or apply less smearing.

\section{Comparison between gradient flow and stout smearing}

\begin{figure}[h]
\centering
\includegraphics[width=0.495\textwidth]{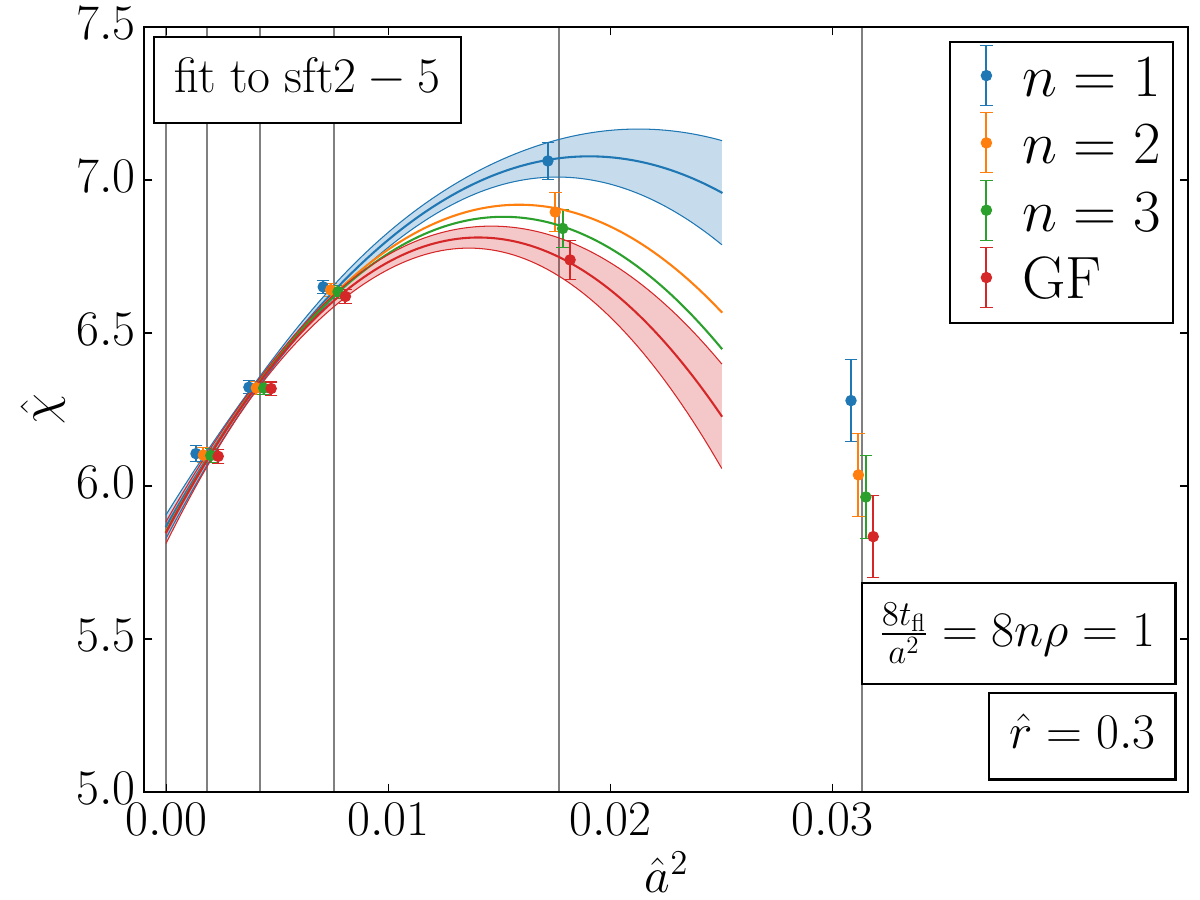}
\includegraphics[width=0.495\textwidth]{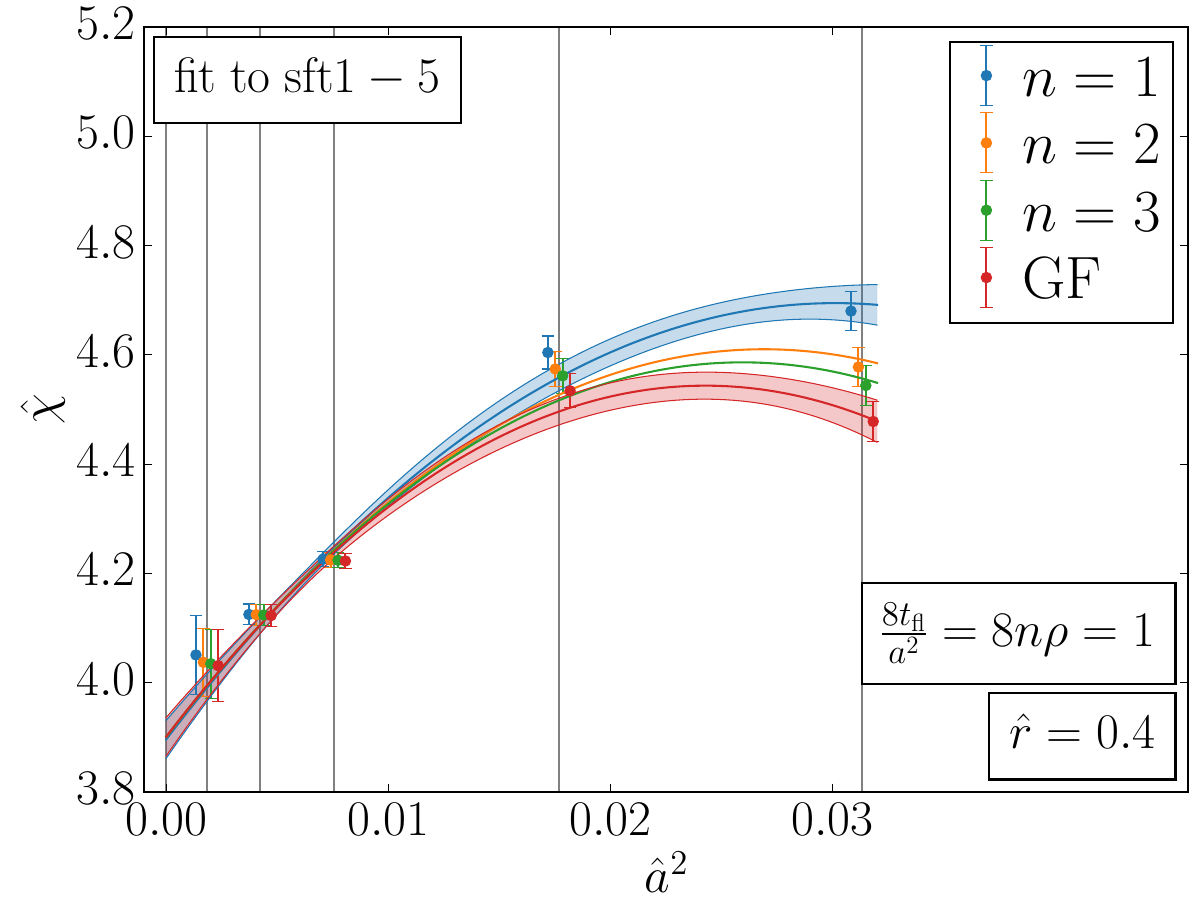}
\vspace{-0.6cm}
\caption{Continuum extrapolations of stout smeared Creutz ratio $\chihat$ at distances $\rhat=0.3/0.4$ ($r=0.14/0.18\,\fm$) as a function of the lattice spacing $\ahat^{2}$. Data points have been shifted for better visibility.}
\label{fig:continuumstout}
\end{figure}
In simulations including dynamical fermions the application of the gradient flow as a gauge field smearing technique in the Dirac operator is computationally demanding. Instead, one therefore employs stout smearing with a small number of smearing iterations. However, gradient flow and stout smearing can be related to each other by $\frac{8 \tflow}{a^{2}}=8n\rho$, where $\rho$ is the stout smearing parameter, as in the limit of $n\to\infty$ and $\rho \to 0$ stout smearing converges towards the gradient flow~\cite{Nagatsuka:2023jos}, i.e. stout smearing can be understood as an approximation of gradient flow smearing. In \cref{fig:continuumstout} continuum extrapolations for stout and gradient flow smeared Creutz ratios are displayed with a smearing strength of $\frac{8 \tflow}{a^{2}}=8n\rho=1$. We observe that replacing the gradient flow by stout smearing increases the absolute size of lattice artefacts. However, when applying three stout iterations, the stout smeared Creutz ratio almost reproduces the gradient flow result within errors. The location of the maximum of the stout smeared extrapolation is shifted to somewhat larger $\ahat^{2}$ compared to the gradient flow, i.e. our findings for gradient flow smearing form a conservative bound. In principle, even the application of one stout smearing iteration might be sufficient.

\section{Conclusion and Outlook}

We have investigated the influence of gradient flow smearing and physical gradient flow on continuum extrapolations. In particular, we studied diagonal Creutz ratios $\chi(r,r)$ evaluated at various distances $r$. We introduced an extrapolation ansatz that allows to simultaneously study gradient flow smearing and physical gradient flow. Our investigations show that the maximum tolerable smearing radius still allowing for a monotonous continuum extrapolation depends on the distance $r$. We demand monotony as a minimum requirement for a reliable extrapolation of data to the continuum as we only have limited knowledge about the shape of the continuum extrapolation, in particular when higher order $a$ effects and logarithmic corrections~\cite{Husung:2017qjz} have to be considered, i.e. a simple functional form is essential for a controlled extrapolation. Given a set of lattice gauge ensembles covering lattice spacings in a certain fixed range, we confirmed that for short distance observables less smearing is tolerable. \Cref{fig:peaklocation} summerises the main result of our numerical investigation. Each curve yields the upper bound of the region where a continuum extrapolation is monotonous. For $\frac{8\tflow}{a^{2}}= 1$ we also studied the case when gradient flow smearing is approximated by a small number $n$ of stout smearing iterations keeping $\frac{8 \tflow}{a^{2}}=8n\rho$ fixed. In this case we found that the location of the maximum in the continuum extrapolation is shifted to somewhat larger $\ahat^{2}$. Moreover, even a single stout smearing step with $\rho=\frac{1}{8}$ reproduces our conceptual findings. Creutz ratios give a measure of the force between static quarks and hence also have an implication for computations including fermions. Nevertheless, it is important to confirm our findings also considering observables based on fermions, where the smearing strength is fixed to the found range.

\vspace{1em}
\begin{small}
We gratefully acknowledge the support of DESY where our computations were performed. AR would like to thank S. Schaefer and R. Sommer for a fruitful collaboration.
\end{small}

\bibliographystyle{JHEP}
\bibliography{proceedings.bib}

\end{document}